\documentclass[
aps,%
12pt,%
final,%
notitlepage,%
oneside,%
onecolumn,%
nobibnotes,%
nofootinbib,%
superscriptaddress,%
noshowpacs,%
centertags]%
{revtex4}

\usepackage{graphicx} 

\begin{document}

\title{
Compactification and non-zero density in (2+1)-dimensional four-fermion model}%

\author{\firstname{K. G.}~\surname{Klimenko}}
\affiliation{State Research Center of Russian Federation -- Institute for High Energy Physics,	NRC "Kurchatov Institute", 142281, Protvino, Moscow Region, Russia}%
    
\author{\firstname{ A. N.}~\surname{Kozhakin}}

\affiliation{State Research Center of Russian Federation -- Institute for High Energy Physics,	NRC "Kurchatov Institute", 142281, Protvino, Moscow Region, Russia}%

\author{\firstname{R. N.}~\surname{Zhokhov}}
\email{zhokhovr@gmail.com}

\affiliation{Pushkov Institute of Terrestrial Magnetism, Ionosphere and Radiowave 
Propagation (IZMIRAN), 108840 Troitsk, Moscow, Russia
}%
\affiliation{State Research Center of Russian Federation -- Institute for High Energy Physics,	NRC "Kurchatov Institute", 142281, Protvino, Moscow Region, Russia}%


\begin{abstract}
The phase diagram of (2+1)-dimensional NJL type model 
has been studied
with one compact dimension. The effect of compactification and fermion density on the chiral symmetry breaking is considered for periodic boundary conditions.
\end{abstract}

\maketitle

The NJL model has been investigated intensively as effective model for QCD \cite{Buballa:2003qv}.  
For physical scenarios such as, for example, heavy-ion collisions there are always effects of finite system size and one in principle should account for finite size effects in field theoretical calculations performed as a rule in the thermodynamic limit. One can expect that a sufficiently small size of the system can affect the phase diagram. Finite size effects and compactification have been also studied in (3+1)-dimensional NJL model \cite{Abreu:2019czp, Inagaki:2021yhi, Inagaki:2019kbc, Xu:2019gia, Ebert:2010eq, Abreu:2020uxc, Abreu:2021btt, Correa:2023nmf, Singh:2025nri, Kiriyama:2002xy, Bhattacharyya:2014uxa}

 In low dimensions NJL type models 
 have also received considerable attention since they could be viewed as toy models and they share many properties with QCD \cite{Thies:2006ti, Thies:2018qgx, Thies:2026tth, Thies2, thies, Thies:2025mro, kkz, Ebert:2008us}.
These models could be also used in description of condensed matter systems \cite{Chodos:1993mf, Thies:2006ti, Thies:2026tth, Thies:2025mro, Gusynin:2007ix, Caldas:2009zz, Gomes:2023vvu, Caldas:2024kuu, Caldas:2008xa, Gomes:2021nem, Caldas:2010qi, Caldas:2011fm, Gomes:2022dmf, Lenz:2023wvk, Lenz:2023gsq, Klimenko:2013gua}.

We tried to investigate inhomogeneous phases in compactified  NJL model and found that, although, (2+1)-dimensional NJL model was studied with compact dimension in \cite{Bietenholz:2003wa, Khanna:2012zz, Zhukovsky:2023jsw, Zhukovsky:2013ima, Zhukovsky:2021qik, Zhokhov:2015tga, Ebert:2015hva, Zhokhov:2016xzp,Ebert:2017udh, Ebert:2015vua}, the fermion density in the case of compactified dimension has not been thoroughly considered. In \cite{Zhokhov:2015tga, Ebert:2015hva} only ($T$, $\mu$) phase diagram was considered for only one value of $L$, namely $L=2\ln2/(\pi|g|)$, it corresponds to rather small $L/|g|<1$. Or even more general case with Zeeman effect was considered in \cite{Zhokhov:2016xzp} but without consideration of fermion density.  
In this paper the behavior of the phase structure with varying values of circumference of compact dimension $L$ is investigated and the phase diagram ($L$, $\mu$) is presented.

This type of models is interesting as in the general field theory description (just for example we have not found it when needed) as well as possible applications to condensed matter systems such as graphene, in particular, compactification is relevant to such systems as carbon nano-tubes and nano-ribbons \cite{Bunder, SantosChico}.

In this paper NJL model 
is considered at non-zero fermion density chemical potential in (2+1)-dimensional space-time where one of the spatial coordinates is compactified.

We consider (2+1)-dimensional NJL model
\begin{equation}
L=\bar \psi\Big [\gamma^\rho\mathrm{i}\partial_\rho
+\mu\gamma^0\Big ]\psi+ \frac {G}{N}\Big
[(\bar \psi\psi)^2+(\bar \psi\mathrm{i}\gamma^5 \psi)^2 \Big ],
\end{equation}
where there is one fermion field $\psi(t,\vec x)$ that lies in a reducible spinor representation of the (2+1)-dimensional Lorentz group.
In this reducible four-dimensional representation of Lorentz group $\gamma$-matrices are (4$\times$4) and have the form $\gamma^\mu=diag(\tilde\gamma^\mu,-\tilde\gamma^\mu)$
where $2\times 2$ dimensional $\tilde\gamma$-matrices are defined as
$\tilde\gamma^0=\sigma_3,\,\,
\tilde\gamma^1=i\sigma_1,\,\,
\tilde\gamma^2=i\sigma_2$. $\gamma^5$ matrix, which anticommutes
with $\gamma^\mu~~(\mu=0,1,2)$, is defined by
$\gamma^5={\tiny
i\left (\begin{array}{cc}
0~,& -I\\
I~,& 0
\end{array}\right)}$.

The fermion field is also transformed under the $O(N)$ group. It is introduced just to consider the model in the leading order of the large $N$-expansion. 
$\mu$ is the fermion density chemical potential.

Obviously, the Lagrangian of the model is invariant with respect to $U_V(1)\times U_A(1)$ groups, where $U_V(1)$: $\psi\to e^{i\alpha}\psi$, and continuous chiral (axial)  transformations $U_A(1)$: $\psi\to e^{i\alpha\gamma^5}\psi$.

It is convenient to use the version of the Lagrangian, which is linearized and possesses two auxiliary bosonic fields: the scalar field $\sigma (t,\vec x)$ and the pseudoscalar one $\pi (t,\vec x)$)
\begin{eqnarray}
\widetilde L =\bar \psi\Big [\gamma^\rho\mathrm{i}\partial_\rho
+\mu\gamma^0-\sigma -\mathrm{i}\gamma^5\pi\Big ]\psi
 -\frac{N}{4G}\Big [\sigma^2+\pi^2\Big ].
\label{2}
\end{eqnarray}
One can easily see that the Lagrangians are equivalent to each other, this fact can be demonstrated from bosonic fields equations of motion
\begin{eqnarray}
\sigma(t,\vec x)=-2\frac G{N}(\bar \psi\psi);~~~\pi (t,\vec x)=-2\frac G{N}(\bar \psi
\mathrm{i}\gamma^5 \psi). \label{200}
\end{eqnarray}
In order to study the phase structure of the model  one needs to calculate the thermodynamic potential (TDP)
that can be defined by
\begin{equation}
\int dt d^2\vec x\, \Omega (M, \pi)=-\frac{1}{N}{\cal S}_{\rm {eff}}\{\sigma(t,\vec x),\pi(t,\vec x)\}\big|_{\sigma (t,\vec x)=\langle\sigma\rangle,\,\pi(t,\vec x)=\langle\pi\rangle},
\end{equation}
where ${\cal S}_{\rm {eff}}(\sigma,\pi)$ is effective action that can be obtained in the leading order of the large $N$-expansion.

We suppose that the ground state is homogeneous and condensates do not depend on spatial coordinates $\vec x$. And we assume the following ansatz without pion condensation
\begin{eqnarray}
\langle\sigma(t,\vec x)\rangle=M,~~~\langle\pi(t,\vec x)\rangle=0, \label{6}
\end{eqnarray}
In the paper we planned to study the phase structure in the case of compactified spatial dimensions. Let us assume that the second spatial coordinate is compactified. Then in compactified case one has to impose some boundary conditions and we suppose that  fermion fields $\psi$ satisfy periodic boundary condition of the form
\begin{eqnarray}
\psi(t,x_1,x_2+L)=\psi(t,x_1,x_2),
\label{0t}
\end{eqnarray}
where $L$ is the length of the circumference of compactified dimension.

For compactified graphene sheets boundary conditions for   
the fermion field $\psi(t,x)=(\psi_1(t,x),\psi_2(t,x))^T$ (where $\psi_1$, $\psi_2$ spinors which lies in irreducible representation) have the following form
\begin{eqnarray}
\psi_1(t,x_1,x_2+L)=e^{2\pi i(\nu/3)}\psi_1(t,x_1,x_2),\;\;\;\;\;\;\;\;\;\psi_2(t,x_1,x_2+L)=e^{2\pi i(-\nu/3)}\psi_2(t,x_1,x_2),
\end{eqnarray}
where $\nu$ could be equal to $0$ and $1$ \cite{Ebert:2015hva, Ando, AndoVietAjki}. In the present paper we limit ourselves to the simplest case of $\nu=0$, the case in which the nanotube
exhibits metallic behaviour and we study the possibility of generation of the gap $M$ in the spectrum.

In order to obtain the TDP, one needs to replace the integration over $p_2$-momentum into summation according to the following rule
\begin{eqnarray}
\int_{-\infty}^\infty\frac{dp_2}{2\pi}f(p_2)\longrightarrow \frac 1L\sum_{n=-\infty}^\infty 
f(p_{2n}),~~p_{2n}=\frac{2\pi n}L ,~~ ~~n=0,\pm 1,\pm 2,....
\label{0t2}
\end{eqnarray}
After somewhat tedious calculations one can show that the TDP of the model can be written in the form
\begin{eqnarray}
\Omega(M)=V(M)+\widetilde V_{L}(M)+W_{\mu L}(M)
\end{eqnarray}
where $V(M)+\widetilde V_{L}(M)$ represents the TDP at zero density, i. e. at $\mu=0$, but with compactified dimension, at $L\ne\infty$. $V(M)$ is the TDP of the model at zero $\mu$ and $L=\infty$.
$\widetilde V_{L}(M)$ is the part that is contributed by compactification and can be shown to be equal to
\begin{eqnarray}
\widetilde V_L(M)=-
\frac 4L\int_{-\infty}^\infty\frac{dp_1}{2\pi}\int_{-\infty}^\infty \frac{dp_3}{2\pi}\ln\Big[1-e^{-LE}\Big], \;\;\;\;\text{where}\;\; \;E=\sqrt{p_3^2+p_1^2+M^2}
\label{0t20}
\end{eqnarray}
The term $W_{\mu L}(M)$ is term that depends on the fermion density (if $\mu=0$, then the term $W_{\mu L}(M)$ is equal to zero) and has the following form
\begin{eqnarray}
W_{\mu L}(M)=-
\frac 2L\int_{-\infty}^\infty\frac{dp_1}{2\pi}\sum_{l=-\infty}^\infty\Bigg(\mu-\sqrt{M^2+p_1^2+
\frac{4\pi^2l^2}{L^2}}\Bigg)\theta\Bigg(\mu-\sqrt{M^2+p_1^2+
\frac{4\pi^2l^2}{L^2}}\Bigg)
\label{0t21}
\end{eqnarray}

The obtained TDP is unrenormalized, it means that it is UV divergent. Actually only $V(M)$, i. e. the TDP of the system at zero density and without compactification, $\mu=0$ and $L=\infty$, possesses the divergences and other parts are finite. To obtain a finite expression one needs to regularize and then renormalize it.
We regularize the integral by cutting
momenta and assume that $|p_1|<\Lambda$, $|p_2|<\Lambda$ and then renormalize it using the following expression 
\begin{eqnarray}
\frac
1{4G(\Lambda)}=\frac{2\Lambda\ln(1+\sqrt{2})}{\pi^2}+\frac{1}{2g},
\label{16}
\end{eqnarray}
where model parameter $g$ does not depend on $\Lambda$ and, moreover, it is a finite quantity. So there is one dimensional parameter $|g|$ in the model and we express all the quantities in it, such as $|g|M$, $L/|g|$ and $|g|\mu$  etc. Then,
one gets the following renormalized expression for the effective potential $V(M)$, i.e.
\begin{eqnarray}
V(M)=V^{ren}(M)
=\frac{M^2}{2g}+\frac{M^3}{3\pi}. \label{17}
\end{eqnarray}
It is evident that if $g>0$ then the global minimum of the TDP lies at $M=0$ and chiral symmetry is not broken. 
However, in the case if it is negative, i. e. $g<0$, then the TDP has
global minimum at the value
$M_0=-\pi/g$ and dynamical fermion mass $M_0$ is generated and chiral symmetry gets broken down.

\vspace{30pt}

The phase diagram of NJL model at $g<0$ without compactification, i. e. $L=\infty$, is well known. In vacuum, i. e. at zero density, $\mu=0$, and zero temperature (as in our case), the chiral symmetry of the system is broken down dynamically and fermion has non-zero mass $M_0=-\pi/g$. If $\mu$ is larger than fermion mass $M_0$ then there is a first order phase transition from the phase with chiral symmetry breaking to the phase with restored chiral symmetry and zero fermion mass, this phase is called symmetric phase. 

The phase diagram with finite values of $L$ has been explored and one can contemplate the obtained ($L$, $\mu$)-phase diagram in Fig. 1. One can see that in the limit $L\to\infty$ the phase structure is tending to the one that was described above for the case without compactification.

With compactification the picture appeared to be rather complicated, let us discuss it. Note that there is a pronounced phase transition from chiral symmetry broken (CSB) phase to symmetric one (SYM) that is rather smooth (ignore the peaks for a moment), it goes from large $L$ where $\mu=M_0$ and first the value of $\mu_c$, at which the phase transition takes place, decreases, it means that for fermion density it is easier to restore the chiral symmetry at finite values of $L$. It drops considerably for small circumferences $L$ but then increases substantially for very small $L$, so that it is rather hard if possible for $\mu$ to restore the symmetry. So finite values of $L$ first disfavour the chiral symmetry breaking and further at even smaller values it catalyzes it quite substantially. So the effect of finite compact dimension is not quite simple. Additionally, one can notice that there are several first order phase transitions between CSB phases and there exist non-trivial horns at some values of circumference length. So for some values of $L$ it is harder for $\mu$ to restore the symmetry than for other values around. One can also note that if one decreases the value of $L$ for some fixed values of $\mu$ one can go through the series of phase transitions from broken to restored symmetry and back. It can be interesting, for example, to condensed matter systems applications. Let us discuss the phase transitions in the figure, along $\mu$ at $L=1.5/|g|$ first there is a jump in condensate $M$ from bulk CSB phase to "horn" part (so the phase transition of the first order), then there is a second order phase transition into symmetric phase. At values of $L$ larger than the maximum of the "horn", for example at $L=2.5/|g|$ the first order phase transition gets weaker but the transition into symmetric phase is of first order. The same picture take place around other "horns". Also at small values of $L$ the phase transition into symmetric phase is of first order. 

\begin{figure}
\includegraphics[width=0.7\textwidth]{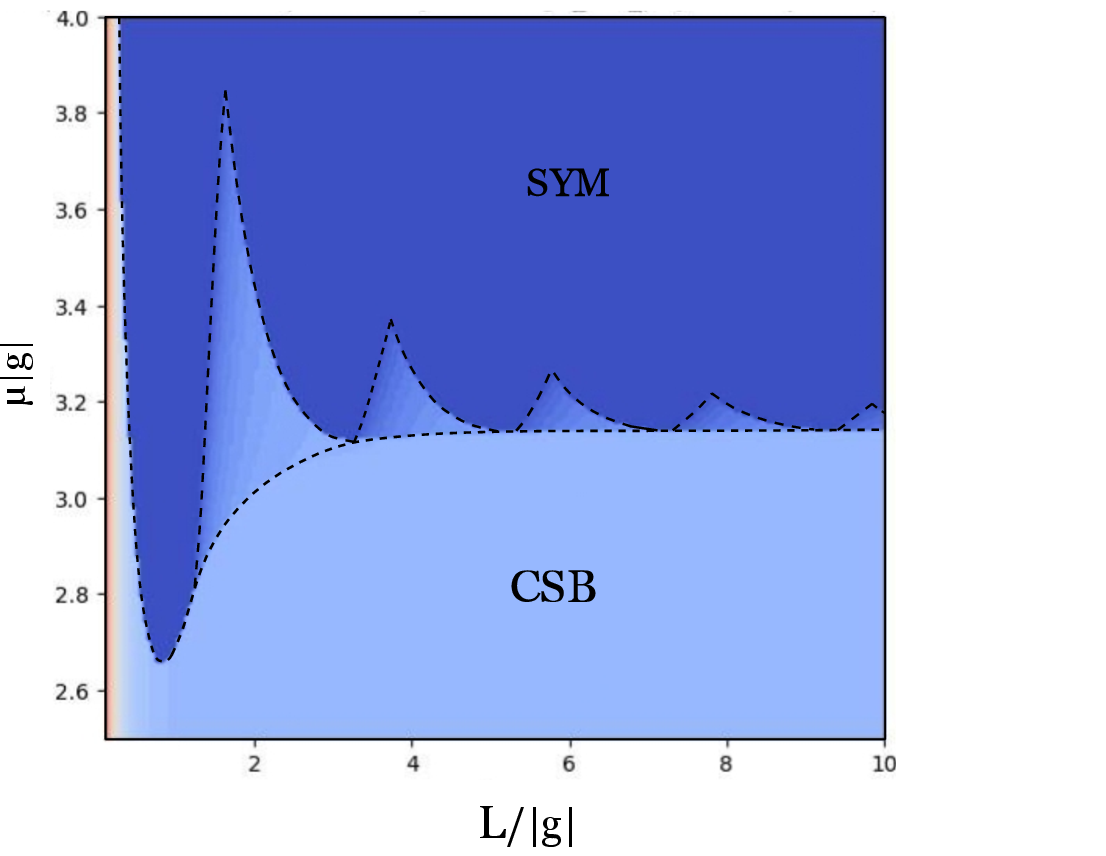}
\caption{($L$, $\mu$)-phase diagram  at g=-1. CSB phase is chiral symmetry broken phase where $M\neq0$, SYM is symmetric phase where $M=0$.}
\end{figure}

\section*{Conclusion}
The phase diagram of (2+1)-dimensional NJL type model 
has been investigated
with compactification (with one compact dimension) at non-zero fermion number chemical potential (fermion density). Periodic boundary condition (for fermions) for compact dimension was considered. The non-trivial effect of compact dimension on chiral symmetry breaking and dynamical generation of fermion mass is observed. In compact system chiral symmetry could be broken even though in the limit of large dimension it was restored by large fermion density, and vice versa it can be restored even if it was broken in the limit of $L\to\infty$. And there could be a series of phase transitions by altering the length of circumference of compactified dimension. These models are interesting as a general field theory considerations and for applications to rolled planar systems in condensed matter physics such as graphene tubes and ribbons.

\end{document}